\documentclass[twocolumn,superscriptaddress,aps,pre]{revtex4-2}

\usepackage{url}
\usepackage{bm}
\usepackage{comment}
\usepackage{amssymb}
\usepackage{amsmath}
\usepackage{graphicx}
\usepackage{color}
\usepackage{empheq}

\begin{document}

\title{Butterfly Effects in Turbulence: A Unified View through the Finite
Size Lyapunov Exponent} 

\author{Victor  de Jesus  Valad\~ao}
\affiliation{Department of Physics and INFN, University of Rome “Tor Vergata”, Via della Ricerca Scientifica 1, 00133 Rome, Italy.}

\author{Erik Aurell}%
\affiliation{Department of Computational Science and Technology, AlbaNova University Center, SE-106 91 Stockholm, Sweden.}%

\author{Guido Boffetta}%
\affiliation{Dipartimento di Fisica and INFN - Universit\`a degli  Studi di Torino, Via Pietro Giuria, 1, 10125 Torino TO, Italy.}%

\author{Massimo Cencini}%
\email{massimo.cenicni@cnr.it}
\affiliation{Istituto dei Sistemi Complessi, CNR Via dei Taurini 19 00185 Rome, Italy.}%
\affiliation{INFN Sezione Roma “Tor Vergata”, Via della Ricerca Scientifica 1, 00133 Rome, Italy.}

\author{Stefano Musacchio}%
\affiliation{Dipartimento di Fisica and INFN - Universit\`a degli  Studi di Torino, Via Pietro Giuria, 1, 10125 Torino TO, Italy.}%

\author{Angelo Vulpiani}%
\affiliation{Istituto dei Sistemi Complessi, CNR Via dei Taurini 19 00185 Rome, Italy.}
\affiliation{Dipartimento di Fisica, Universit\`a di Roma “La Sapienza”, P.le Aldo Moro 5, 00185, Rome, Italy.}%

\date{\today}

\begin{abstract}
  The ``butterfly effect'', introduced over half a century ago by Edward Lorenz, has shifted from a cornerstone of dynamical systems to a popular metaphor, yet its true physical manifestation in fully developed turbulence spans a spectrum of phenomena from standard sensitivity on initial conditions to the recently established concept of Eulerian spontaneous stochasticity. Here, we propose an attempt at a systematic synthesis that brings these different but interconnected ideas together within the unifying framework of the Finite Size Lyapunov Exponent (FSLE). The FSLE describes the growth rate of perturbations as a function of their scale, enabling a comprehensive characterization of the multiscale physics of turbulent flows. Using the FSLE and the Sabra shell model, extended to include thermal noise, we bridge the classical, small-scale Lyapunov regime with predictability at large scales and its interpretation in terms of Eulerian spontaneous stochasticity. Moreover, using the FSLE and the Kraichnan model, we also illustrate the closely related phenomenon of Lagrangian spontaneous stochasticity. To complete the spectrum of butterfly effects, we also examine the ``literal butterfly'' scenario of localized, sub-dissipative perturbations.
\end{abstract}


\maketitle

\section{Introduction\label{sec:intro}}
Nowadays, it is common to ask questions about any subject to generative AI. We asked  \textit{Can you explain in simple terms what the ``butterfly effect" is?} to a couple of them. The answers were broadly similar, providing the core intuition but accompanied by some imprecise or vague statements. Some sentences from both sources are:

\textit{The butterfly effect is a simple idea from a field called Chaos Theory. In plain terms, it means: Small actions can lead to big, unexpected consequences over time}.

\textit{The concept gets its name from a classic example: ``A butterfly flaps its wings in Brazil, and weeks later, those tiny ripples in the air influence weather patterns until a tornado forms in Texas'' Does a butterfly literally create a tornado? Not exactly. It’s more about sensitivity. In a complex system (like the weather, the economy, or your own life), the starting conditions are so sensitive that even the smallest tweak can completely change the final outcome.} \textit{Everyday example: You leave home 2 minutes late; That makes you miss a bus; You meet a different person later; That meeting changes your job or life path.}

Reformulating the question in more precise terms, by inputting prompts that drive the AI toward a mathematical or physical background, the answers become significantly more rigorous. However, the above-reported answers are interesting and insightful: being based on the vast amount of texts ``ingested'' by the AI during its training, they reveal how thoroughly the \textit{butterfly effect} has become popular since Lorenz's seminal work \cite{lorenz1972} (see e.g. Hilborn~\cite{hilborn2004sea}). As emphasized by Ghys \cite{ghys2015butterfly}, it is profoundly ``unusual for a mathematical idea to disseminate into society at large'' quite like Lorenz's butterfly.

In its common parlance, the butterfly effect is an appealing expression to convey the concept of sensitive dependence on initial ($t=0$) conditions of chaotic systems. Typically, it is a synonym for the exponential divergence of infinitesimal perturbations or, in mathematical terms, the presence of at least one positive Lyapunov exponent ($\lambda >0$). Roughly, expressed in formulas, if $\delta_0$ denotes the initial magnitude of the perturbation, it is expected that at time $t$ the perturbation size has grown as
\begin{equation}
\delta(t) \approx \delta_0 \exp(\lambda t)\,. \label{eq:1}
\end{equation}  
Consequently, if a macroscopic prediction tolerates uncertainties up to a given threshold value $\Delta$, the predictability time, $T_p$, scales as
\begin{equation}
T_p \approx \frac{1}{\lambda} \ln\left(\frac{\Delta}{\delta_0}\right)\label{eq:2}\,,
\end{equation}  
which, however, holds if and only if both $\delta_0$ and $\Delta$ are infinitesimal.  At first glance, Eq.~(\ref{eq:2}) implies that $T_p$ can be extended indefinitely, provided the initial uncertainty is made sufficiently small (though the logarithmic dependence severely limits this practical gain, as discussed in the following sections).

In recent years, the butterfly effect has gained renewed interest in the context of turbulent (multiscale) systems, and has been ascribed to a more severe form of unpredictability -- called ``the real butterfly effect'' by Palmer,\cite{palmer2014,palmer2024} then promoted as spontaneous stochasticity \cite{mailybaev2016spontaneous,thalabard2020butterfly,bandak2024spontaneous} -- exhibited by multiscale hydrodynamic systems. Such a strong form of unpredictability was foreseen by Lorenz himself, who in the very abstract of his 1969 paper wrote:\cite{lorenz1969predictability}
\begin{quote}
It is proposed that certain formally deterministic fluid systems which possess many scales of motion are observationally indistinguishable from indeterministic systems; specifically that two states of the system differing initially by a small observational error will evolve into two states differing as greatly as randomly chosen states of the system within a finite time interval, which cannot be lengthened by reducing the amplitude of the initial error.
\end{quote}
The sentence above indeed contradicts the very essence of Eq.~(\ref{eq:2})!

Recently, a sort of classification of butterfly effects was proposed, \cite{shen2022three,pielke2024butterfly} where, besides the more standard definition in terms of sensitive dependence on initial conditions (in the sense of a positive Lyapunov exponent) and the aforementioned  strong form of unpredictability summarized above in Lorenz's words, the literal possibility of  ``the actual flapping of a real butterfly'', in the sense of a small (infinitesimal) spatially localized perturbation, was also considered.\cite{pielke2024butterfly} Finally, the size of these perturbations -- while mathematically arbitrary -- must be tied to well-defined physical quantities when modeling real systems like turbulent flows. As we will see, this implies that, for very small perturbations, thermal noise becomes significant, requiring a more accurate description via the Landau–Lifschitz fluctuating hydrodynamics equations.\cite{bell2022thermal} Thus, the role of the noise must be included in the problem of predictability in turbulence.

In this paper, we revisit the butterfly effects in the context of turbulence. In particular, we employ shell models of turbulence \cite{bohr1998,biferale2003} as a simplified and illustrative framework to investigate phenomena that also arise in real turbulent flows. Furthermore, concerning the aforementioned different butterfly effects, we re-examine them using the finite-size Lyapunov exponent (FSLE) \cite{aurell1996growth,aurell1997predictability,boffetta2002predictability,cencini2013finite}, showing how it bridges classical approaches to predictability in turbulence to the recent discussions on Eulerian spontaneous stochasticity\cite{mailybaev2016spontaneous,thalabard2020butterfly,bandak2024spontaneous}. Spontaneous stochasticity is easier to understand in the Lagrangian context, i.e., when considering the evolution of particle pairs in turbulent flows -- the Richardson dispersion~\cite{richardson1926atmospheric} --, where it was first discovered.\cite{bernard1998slow,gawedzki2000phase,falkovich2001particles} Here, we will reformulate it in the language of the FSLE and then extend this approach to the Eulerian context. Concerning the ``literal'' butterfly effect, shell models do not possess a spatial structure, so the notion of spatially localized perturbations is not well defined. Nevertheless, we can mimic such perturbations by introducing an initial perturbation localized at the dissipative scales and then following its evolution. In chaotic systems, infinitesimal perturbations grow exponentially in time, as described by (\ref{eq:1}), in the asymptotic regime. However, in the transient behavior, they may initially decrease before growing exponentially. We will show that, in the shell model, a perturbation initially localized at the dissipative scales indeed dramatically illustrates the role of such transient dynamics. Moreover, we will also discuss the role of thermal fluctuations in limiting the initial decrease of such perturbations.

The paper is organized as follows. In Sec.~\ref{sec:shellmodel}, we briefly introduce the Sabra shell model~\cite{procaccia} for turbulence and its stochastic variant, which mimics the effect of thermal noise~\cite{bandak}.  In Sec.~\ref{sec:pred}, after reviewing classical arguments for predictability in chaotic multiscale systems, we show how these foundational concepts map directly onto the behavior of the Finite Size Lyapunov Exponent. We then end the section discussing the literal butterfly effect. In Sec.~\ref{sec:spontaneous}, reframing the framework established in Ref.~\cite{bandak2024spontaneous}, we discuss how the FSLE can be used to characterize the phenomenon of Eulerian spontaneous stochasticity. In doing so, we will first open a digression on Lagrangian spontaneous stochasticity, concerning the dynamics of tracers in turbulent flows, with an illustrative example based on the Kraichnan model~\cite{kraichnan1968small,bernard1998slow,gawedzki2000phase,falkovich2001particles}. Section~\ref{sec:discussions} is devoted to conclusions and discussion. In our treatment, for the sake of simplicity, we neglect the effects of intermittency; these are briefly considered in Appendix~\ref{app:fsle}, where we also provide some technical details on the computation of the FSLE.

%%%%%%%%%%%%%%%%%%%%%%%%%%%%%%%%%%%%%%%%%%%%%%%%%%%%%%%%%%%%%%%%%%%%%%%%%%  
\section{Shell models for turbulence\label{sec:shellmodel}}
%%%%%%%%%%%%%%%%%%%%%%%%%%%%%%%%%%%%%%%%%%%%%%%%%%%%%%%%%%%%%%%%%%%%%%%%%%  
As anticipated, we illustrate the main issues of predictability and the butterfly effect in turbulence using a shell model for turbulence.\cite{bohr1998,biferale2003} Such a class of models mimics the turbulent energy cascade and, while lacking spatial structures, it preserves its multiscale character, which is at the heart of the predictability problem in turbulence. In the following two subsections, we first introduce the (deterministic) Sabra shell model \cite{procaccia} and then its generalization that accounts for the presence of thermal fluctuations.\cite{bandak}

%%\\\\\\\\\\\\\\\\\\\\\\\\\\\\\\\\\\\\\\\\\\\\\\\\\\\\\\\\\\\\\\\\\\\\
\subsection{Sabra shell model}
%%\\\\\\\\\\\\\\\\\\\\\\\\\\\\\\\\\\\\\\\\\\\\\\\\\\\\\\\\\\\\\\\\\\\\
In shell models, scales are discretized in $N$ wavenumbers $k_n = k_0 2^{n-1}$ ($n = 1,\ldots,N$) describing a sequence of logarithmically equispaced, spherical shells in k-space. For each shell there is an associated complex variable $u_n$, representing the velocity fluctuations at scale $\ell_n=1/k_n$, evolving according to the dynamics \cite{procaccia}:
\begin{eqnarray}
  &&  \frac{du_n}{dt}= i \left( k_{n+1}u_{n+1}u^{*}_{n+1}-\frac{1}{2} k_{n}u_{n+1}u^{*}_{n-1} +\right. \label{eq:Sabra} \\
&&  \left. \frac{1}{2} k_{n-1} u_{n-1}u_{n-2}\right)-\nu k_n^2 u_n+f_n\,,  \nonumber
\end{eqnarray}  
where  $^*$ denotes complex conjugation, and boundary conditions $u_{m}=0$ for $m=-1,0$ and $m=N+1,N+2$ are assumed. Equation~(\ref{eq:Sabra}) mimics the Navier-Stokes equation in Fourier space: the quadratic term corresponds to advection, the viscosity $\nu$ controls dissipation, while the forcing $f_n$ injects energy at an average rate $\epsilon=\langle \sum_n \Re(f_n u_n^{*})\rangle$ (brackets represent time average),  typically only at the largest scales (small $n$). In the unforced, inviscid limit ($f_n=\nu=0$), the nonlinear term preserves energy $E$ and helicity $H$,
\begin{equation}
E=\frac{1}{2}\sum_{n=1}^{N} |u_n|^2 \,,\qquad H=\sum_{n=1}^{N} (-1)^n |u_n|^2\,,
\end{equation}  
as the 3D Euler equation. When forced, $f_n\neq 0$, the nonlinear term induces an energy cascade from the forced scales to the smaller ones, the viscous term becomes active in dissipating energy at the Kolmogorov wavenumber $k_{n_\eta}=(\epsilon/\nu^3)^{1/4}$, which can be readily derived by using Kolmogorov 1941 (K41) type dimensional arguments \cite{frisch95} (see Sec.~\ref{sec:scales}). Far from the forcing and dissipative scale, moments of the velocity variables are characterized by a power law behavior $\langle |u_n|^p\rangle \sim k_n^{-\zeta_p}$ with anomalous exponents $\zeta_p$ deviating from the K41 expectation $p/3$, and quantitatively close to the exponents ruling the behavior of structure functions in Navier-Stokes turbulence.

Therefore, despite their simplicity, shell models display a rich phenomenology that closely follows that of Navier-Stokes turbulence with many characteristic scales (from $k_{n_f}$ to $k_{n_\eta}$) and  times
\begin{equation}
\tau_n=(k_n^2|u_n|^2)^{-1/2}\approx k_n^{-2/3}\,,
\label{eq:ett}
\end{equation}
where, for the sake of simplicity, we neglected the effects induced by intermittency.

 \begin{figure}[t!]
   \includegraphics[width=0.5\textwidth]{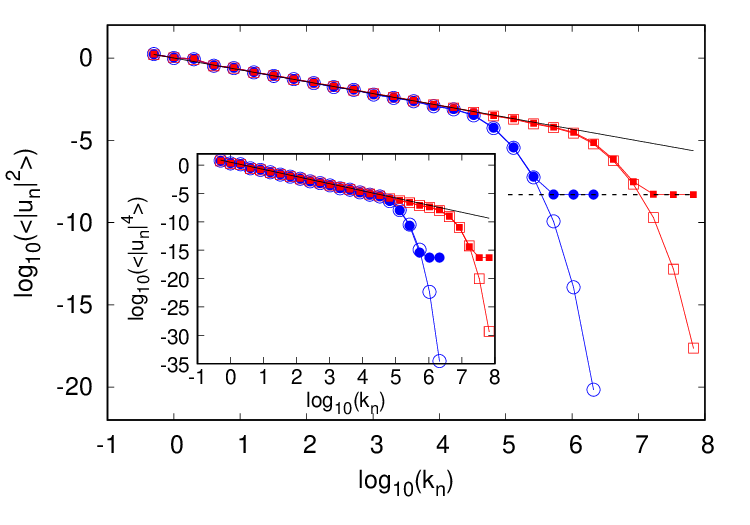}
 \caption{Shell velocity moments in the deterministic and noisy Sabra model: second-order moment for $\nu=10^{-6}$ (blue) and $\nu=10^{-8}$ (red) without (empty symbols) and with (full symbols) thermal noise. The dashed line shows the expected equipartition value $\sigma^2=5\cdot 10^{-9}$ in both simulations. The full black line shows the inertial range power-law exponent $\zeta_2=0.72(1)$. The inset shows the same data for the fourth moment, where $\zeta_4=1.26(1)$. The values of $\zeta_p$'s are unaltered by the presence of noise and match the known values~\cite{procaccia}.  \label{fig:moments}}
 \end{figure}
%%\\\\\\\\\\\\\\\\\\\\\\\\\\\\\\\\\\\\\\\\\\\\\\\\\\\\\\\\\\\\\\\\\\\\
\subsection{Thermal noise and fluctuating hydrodynamics in the shell model framework\label{sec:noisySabra}}
 %%\\\\\\\\\\\\\\\\\\\\\\\\\\\\\\\\\\\\\\\\\\\\\\\\\\\\\\\\\\\\\\\\\\\\
Following Bandak et al.\cite{bandak}, we also consider a \textit{noisy} variant of the Sabra shell model (\ref{eq:Sabra}) that accounts for fluctuations induced by thermal noise, which becomes relevant below the Kolmogorov dissipative scale. Thermal fluctuations modify the deep dissipative range, inducing an equilibrium spectrum at such small scales, \cite{bandak,bell2022thermal} as foreseen by Betchov.\cite{betchov1957fine} While the Sabra model can be considered a toy model for the incompressible Navier-Stokes equation, its stochastic version is a model of the  fluctuating hydrodynamics equation of Landau and Lifschitz. In its stochastic version, the Sabra model takes the form~\cite{bandak}
\begin{eqnarray}
&&  \frac{du_n}{dt}= i \left( k_{n+1}u_{n+1}u^{*}_{n+1}-\frac{1}{2} k_{n}u_{n+1}u^{*}_{n-1}+\right. \label{eq:Sabra-thermal} \\
  &&\left. \frac{1}{2} k_{n-1} u_{n-1}u_{n-2}\right)-\nu k_n^2 u_n+f_n +\left(2\nu \sigma^2\right)^{1/2} k_n \eta_n\,,  \nonumber
\end{eqnarray}  
where  $\eta$ is a zero mean complex white noise with covariance $\langle \eta_n(t)\eta_{m}(s)\rangle=\delta_{nm}\delta(t-s)$. In the context of fluctuating hydrodynamics, the noise strength is controlled by the temperature, and we can write $\sigma^2=k_BT/\rho$, where $k_B$ is the Boltzmann constant and $\rho$ the fluid mass; the noise strength is fixed by the fluctuation-dissipation theorem \cite{bandak}.

The presence of the noisy term in Eq.~(\ref{eq:Sabra-thermal}) leaves the scaling properties of the (deterministic) shell model unmodified in the inertial range, while it induces energy equipartition, i.e. $\langle |u_n|^2 \rangle \approx \sigma^2$, for sufficiently small scales as originally found in Ref.~\cite{bandak} and shown in Fig.~\ref{fig:moments}.  In real 3D turbulence, this means that at sufficiently large wave numbers the energy spectrum behaves as $\sim k^{2}$ as first argued in Ref.~\cite{betchov1957fine} and numerically demonstrated in Ref.\cite{bell2022thermal}. It should be stressed, as also discussed later, that despite the  noise irrelevance in the statistics of scales within the inertial range, it impacts the predictability of large scales as well, as first realized by Ruelle~\cite{ruelle1979microscopic}.

For simulating Eqs.~(\ref{eq:Sabra}) and (\ref{eq:Sabra-thermal}), we used a stochastic 2nd-order Runge-Kutta (Heun) algorithm \cite{honeycutt1992stochastic} with exact integration of the linear terms via integrating factors. We considered two values of viscosity $\nu=10^{-6}$ ($10^{-8}$) with $N=23$ ($28$) shells and time step $dt=10^{-6}$ ($10^{-7}$) and fixed $k_0=1/2$. The forcing is chosen to be a Gaussian white noise acting on the first three shells with amplitude such as to fix the energy input to unity ($\epsilon=1$). For the deterministic case, we checked that the moments of the velocities match those obtained by using a 4th-order Runge-Kutta algorithm.

%%%%%%%%%%%%%%%%%%%%%%%%%%%%%%%%%%%%%%%%%%%%%%%%%%%%%%%%%%%%%%%%%%%%%%%%%%%%%%%
\section{Predictability in turbulence\label{sec:pred}}
%%%%%%%%%%%%%%%%%%%%%%%%%%%%%%%%%%%%%%%%%%%%%%%%%%%%%%%%%%%%%%%%%%%%%%%%%%%%%%%
In this section, after recapitulating the hierarchy of scales characterizing turbulent flows (for shell models they are the same), we  review the classical approaches to predictability in turbulence. We then reframe these classical approaches in terms of the finite-size Lyapunov exponent. We end the section with a brief discussion of the literal butterfly effect mentioned in the introduction.

%///////////////////////////////////////////////////////////////////////
\subsection{Characteristic scales, times and velocities in turbulence\label{sec:scales}}
%///////////////////////////////////////////////////////////////////////
In turbulence, the rate of energy input $\epsilon$, which at stationarity equals the energy dissipation rate, can also be written as $\epsilon=U_0^2/T_0=U_0^3/L_0$, where $U_0$, $L_0$ and $T_0=L_0/U_0$ are the characteristic velocity, length and (eddy turnover) time at scales where the forcing is acting. The corresponding quantities at the small scales where dissipation takes place can be estimated using dimensional arguments by assuming that they only depend on   $\epsilon$ and the viscosity $\nu$:\cite{frisch95} 
\begin{alignat}{2}
u_\eta
&= (\epsilon\nu)^{1/4}
&&= U_0 Re^{-1/4}\,,
\label{eq:kolmo-u}\\
\eta
&= (\nu^3/\epsilon)^{1/4}
&&= L_0 Re^{-3/4}\,,
\label{eq:kolmo-l}\\
\tau_\eta
&= (\nu/\epsilon)^{1/2}
&&= T_0 Re^{-1/2}\,.
\label{eq:kolmo-t}
\end{alignat}
The above quantities  are the so-called Kolmogorov characteristic velocity, length, and time, where $Re=U_0L_0/\nu$ is the Reynolds number. Again using Kolmogorov ideas,\cite{frisch95} eddies of size $\ell$, with $\eta < \ell < L_0$ -- in the inertial range --, should only depend on $\ell$ and $\epsilon$ and thus  are characterized by typical velocity fluctuations $u_\ell \sim (\epsilon \ell)^{1/3}=U_0 (\ell/L_0)^{1/3}$, and the eddy turnover time $\tau_\ell = \ell/u_\ell$ can be written as
\begin{equation}
\tau_\ell=  T_0 \left(\frac{\ell}{L_0}\right)^{2/3}=T_0 \left(\frac{u_\ell}{U_0}\right)^{2}\,,
\label{eq:tempivel}
\end{equation}  
where the last equality directly links the eddy turnover time to the velocity fluctuations.

These dimensional arguments, which are briefly discussed in Appendix~\ref{app:fsle}), show the vast range of characteristic velocities, lengths and times of turbulent flows. We also note that the same arguments can be used for the shell model, and one simply needs to link the length scales to wave numbers as $\ell_n=1/k_n$ with $L_0=1/k_0$. Equipped with the above dimensional estimates, it is now useful to re-examine Eqs.~(\ref{eq:1}) and (\ref{eq:2}).

\subsection{The different mechanisms of perturbation growth: classical considerations\label{sec:classical}}
The mathematical definition of the (maximal) Lyapunov exponent requires two limits in a specific order, namely
\begin{equation}
\lambda=\lim_{t\to \infty} \lim_{\delta_0\to 0} \frac{1}{t} \ln\left(\frac{\delta(t)}{\delta_0}\right)\,.\label{eq:lyap}
\end{equation}  
The  $t\to \infty$ limit is essentially a time average and, by assuming ergodicity, gives a number -- the Lyapunov exponent -- that does not depend on the initial condition, i.e. on the specific point in phase space where the perturbation was initialized \cite{cencini2010}. Mathematically speaking, the limit $\delta_0\to 0$ means that one has to work in tangent space, and thus that for Eq.~(\ref{eq:2}) to hold both $\delta_0$ and $\Delta$ should be infinitesimal, as already said in the introduction. Physically speaking, we are used to saying that both $\delta_0$ and $\Delta$ should be small enough.  But what does it mean ``small enough''?

For low-dimensional systems, with just a single characteristic scale and time -- such as, e.g., the celebrated Lorenz 1963 model~\cite{lorenz1963} -- the meaning of ``small enough'' is rather clear: simply for the exponential regime (\ref{eq:1}) to hold, $\delta(t)$ must remain much smaller than the attractor size. Such a simple scenario does not apply to systems with many characteristic scales and times of motion like turbulence.

In the presence of many time scales, the  Lyapunov exponent is typically associated with the fastest time scale of a system that, in turbulence, is the Kolmogorov time (\ref{eq:kolmo-t}), thus, as first suggested by Ruelle~\cite{ruelle1979microscopic},
\begin{equation}
\lambda \approx \frac{1}{\tau_\eta}=  \frac{1}{T_0} Re^{1/2}\,,
\label{eq:ruelle}
\end{equation}
as approximately confirmed in simulations for shell models.\cite{crisanti1993intermittency} Actually, because of intermittency, corrections to the $1/2$ scaling (\ref{eq:ruelle}) are present; since they are not crucial for the forthcoming discussion, they are briefly discussed in Appendix~\ref{app:fsle}.

What about the size of the perturbation for the Lyapunov exponent to be significant? In turbulence (or in the shell model), the perturbation represents an initial difference in the velocity field (or shell velocities), i.e., $\delta=\delta u$. For the sake of notation simplicity, we will maintain $\delta$ in the following. From the hierarchy of scales described in Sec.~\ref{sec:scales}, it is clear that for Eq.~(\ref{eq:2}) to apply, both $\delta_0$ and $\Delta$ should be smaller than the Kolmogorov velocity (\ref{eq:kolmo-u}). Thus, from a physical point of view, we have to request $\delta_0 < \Delta \lesssim u_\eta=U_0 Re^{-1/4}$. By posing $\Delta=u_\eta$ in the previous expression and using Eq.~(\ref{eq:ruelle}), we thus obtain
\begin{equation}
\delta_0 < U_0 Re^{-1/4} \exp\left(-\frac{T_p}{T_0} Re^{1/2}\right)\,.
\label{eq:delta0}
\end{equation}  
This result shows that in realistic applications with large $Re$ the chaotic regime in which (\ref{eq:2}) can be applied to increase predictability requires unrealistically small initial uncertainties $\delta_0$. One can realize this by considering that a typical (non-extreme) atmospheric wind speed of $10 m s^{-1}$ and a length $L \gtrsim 1 km$ (i.e., the length scale on which turbulence is created by shear) leads to values of $Re\sim O(10^{9})$ and larger. Therefore, in practical applications such as, e.g., the weather forecast, we can only consider $\delta_0$ and tolerances $\Delta$ much larger than the smallest fluctuations, i.e., $u_\eta$. In other words, any conceivable numerical model of the atmosphere will never be able to resolve scales below, or even close to, the Kolmogorov scale (\ref{eq:kolmo-l}), making any reasoning based on the Lyapunov exponent irrelevant.

Therefore, when considering the predictability in turbulence, we must consider how perturbations in the inertial range grow. A classical argument, proposed by Lilly \cite{lilly1972numerical}, is the following. Consider an uncertainty $\delta$, its size will select a scale $k_n$ and heuristically is expected to grow to a larger value, i.e. to the next Fourier shell $n-1$, in a time of the order of the characteristic time of that scale, $\tau_n \propto T_0 (k_n/k_0)^{-2/3}$ as from Eq.~(\ref{eq:ett}). Thus, from shell $n$ to shell $0$, i.e., for the uncertainty to affect the largest scale, it will take a time
\begin{equation}
  T_p= \sum_{i=0}^n \tau_i = T_0 \sum_{i=0}^n \left(\frac{k_i}{k_0}\right)^{-2/3}\,, \label{eq:lilly}
\end{equation}
clearly for the above equation to make sense we should consider Fourier shell reasonably separated e.g. $k_n=\alpha^{n}k_0$ with $\alpha>1$ (for the shell model naturally $\alpha=2$), it easily see that the above series is convergent even in the limit $n\to \infty$ (the physical limit would be $n$ such as $k_n\approx 1/\eta$) provided $\alpha>1$ and that $T_P \approx c T_0$ with $c$ an order one constant.  The essence of the above argument can be understood in terms of a hierarchy of instabilities \cite{Aurell1996a}. Another way to see the result (\ref{eq:lilly}), as also expressed by Lorenz \cite{lorenz1969predictability}, is that there is an inverse cascade of error from the small scales to the large scales.

The physical meaning of this simple reasoning is that the predictability time $T_p$ on the largest scales $L_0=1/k_0$ is \textit{finite}, no matter how small the initial uncertainty is, and it is basically of the order of the characteristic time of the largest scales $T_0$.  It is not difficult to recognize in this result the text quoted in the introduction from Lorenz's 1969 work\cite{lorenz1969predictability}. It is interesting to note that the phenomenological approach by Lilly and Lorenz to the predictability problem was antecedent to the Ruelle result\cite{ruelle1979microscopic}, and without any connection to chaos. The next subsection will reframe the classical predictability problem here summarized in terms of the finite-size Lyapunov exponent.

\subsection{Predictability in terms of the Finite Size Lyapunov Exponent (FSLE)\label{sec:fsle}}

As seen in the previous section, in turbulence, with realistic Reynolds numbers, enhancing predictability by reducing the initial uncertainty, i.e., by exploiting Eq.~(\ref{eq:2}), leads to considering unrealistically small initial uncertainties (\ref{eq:delta0}). Moreover, arguments such as Eq.~(\ref{eq:lilly}) show that Eq.~(\ref{eq:2}) is irrelevant to the predictability problem in turbulence and that the growth of non-infinitesimal perturbations must be considered.  A way to bridge these two regimes of perturbation growth is to introduce the finite-size Lyapunov exponent, \cite{aurell1996growth,aurell1997predictability,cencini2013finite}, which formalizes Lorenz's ideas about the need to go beyond the understanding of very small perturbations when coping with predictability in systems with a multiscale structure.\cite{lorenz1996predictability}

At its core, the FSLE, $\lambda(\delta)$, quantifies the growth rate of a perturbation of size $\delta$, i.e. $\dot{\delta}\approx \delta \lambda(\delta)$.  Though there are different ways to operationally define $\lambda(\delta)$, the basic idea is to compute  the average time $\tau(\delta)$ it takes for an uncertainty $\delta$ to reach for the first time $r \delta$,  (the precise value of $r>1$ is not too important, provided it is not too large) and define
\begin{equation}
\lambda(\delta)= \frac{\ln r}{\langle \tau(\delta)\rangle} \ ,\
\label{eq:fsle-def}
\end{equation}
where the average is taken over many experiments of uncertainty growth (see Refs.~\cite{aurell1997predictability,cencini2013finite} and Appendix~\ref{app:fsle} for further details). One can show that in the limit $\delta \to 0$, $\lambda(\delta)$ recovers the usual Lyapunov exponent while, for larger $\delta$, it describes the nonlinear growth of perturbations, which is in general non-universal and depends on the system under consideration. In low-dimensional systems, with a single characteristic time, it drops to zero for $\delta$ of the order of the attractor size and stays constant $\approx \lambda$ for smaller perturbations. For a review on the applications of the FSLE, see Ref.~\cite{cencini2013finite}

The FSLE, $\lambda(\delta)$, allows us to treat  the predictability for physically relevant uncertainties; in particular instead of  Eq.~(\ref{eq:2}) we can write
\begin{equation}
T_p(\Delta, \delta_0) \sim  \int_{\delta_0}^{\Delta} \frac{d \delta}{\lambda(\delta) \delta}\,,  \label{eq:fsle-tpred}
\end{equation}
which is also valid for non-infinitesimal $\delta_0$ and $\Delta$, with the proviso that the specific value may depend on how the perturbation amplitude is measured.  Indeed, mathematically speaking, unlike the Lyapunov exponent that is independent of the norm used to quantify the perturbation \cite{cencini2010}, $\lambda(\delta)$ may depend on the norm for non-infinitesimal $\delta$. Moreover, the limit $t\to \infty$ cannot be taken when $\delta$ is not infinitesimal, as the perturbation, while growing, will encompass different scales.
\begin{figure}[t!]
\includegraphics[width=0.5\textwidth]{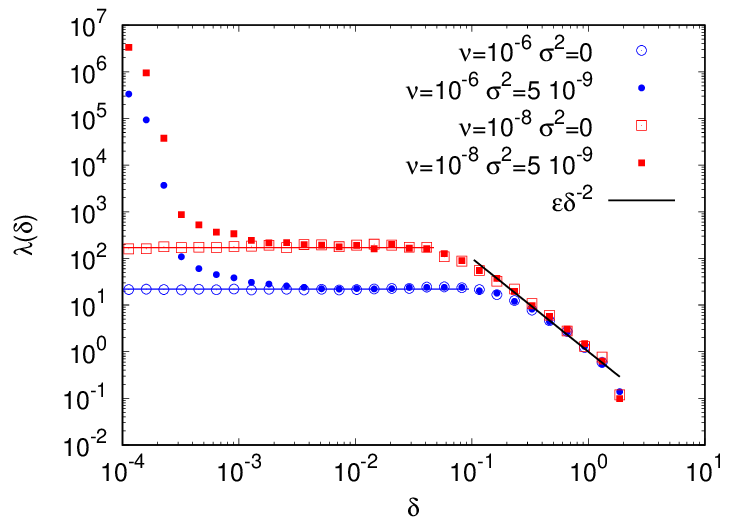}
\caption{Finite size Lyapunov exponent $\lambda(\delta)$ vs $\delta$ in the shell model without (open symbols) and with (filled symbols) thermal noise for two values of the viscosity, $\nu=10^{-6}$ (blue circles) and $10^{-8}$ (red squares), using $N=23$ and $28$ shells, respectively. In the presence of thermal noise, the two replicas evolve with different realizations of the noise and no initial difference in the velocity variables.
 Simulations with thermal noise use $\sigma^2=5\cdot 10^{-9}$ as in Fig.~\ref{fig:moments}. \label{fig:fsle}}
 \end{figure}

In turbulence or, equivalently, in shell models, using simple dimensional arguments from Kolmogorov theory (see Sec.~\ref{sec:scales}) and Ruelle~\cite{ruelle1979microscopic}, we can derive the expected behavior of the FSLE:
\begin{equation}
  \lambda(\delta) \simeq
  \left\{
  \begin{array}{lcl}
\lambda \approx \frac{1}{T_0} Re^{1/2} &\text{if} & \delta \ll u_{\eta}\\
\frac{U_0^2}{T_0}\frac{1}{\delta^2}=\frac{\epsilon}{\delta^2}  &\text{if} & u_{\eta}\ll \delta \ll  U_0 \ .
  \end{array}
  \right.
  \label{eq:fsle}
\end{equation}
For $\delta\gtrsim U_0$ we expect $\lambda(\delta)$ to decrease rapidly as the attractor size is reached.  The first equation is exactly Ruelle's result (\ref{eq:ruelle}) and the second equation can be derived from Eq.~(\ref{eq:tempivel}) by posing $\delta=u_\ell$ and recalling Eq.~(\ref{eq:fsle-def}). In principle, intermittency corrections may appear to the $-2$ scaling exponent, but as shown in Refs.~\cite{aurell1996growth,aurell1997predictability} this is not the case (see also Appendix~\ref{app:fsle}). The two regimes in Eq.~(\ref{eq:fsle}) are demonstrated (open symbols) in Fig.~\ref{fig:fsle} for the Sabra shell model (\ref{eq:Sabra}) for two values of the viscosity (i.e. two different Reynolds numbers), confirming previous numerical results obtained in other shell models \cite{aurell1996growth,aurell1997predictability},  in $3D$-turbulence \cite{boffetta2017chaos} and also in $2D$-turbulence in the regime of inverse energy cascade \cite{boffetta2001}. Figure~\ref{fig:fsle} clearly shows that for $\delta>u_\eta$  the behavior $\lambda(\delta) \approx \epsilon \delta^{-2}$ is independent of the Reynolds number.

Now by using Eqs.~(\ref{eq:fsle-tpred}) and  (\ref{eq:fsle}) to estimate the predictability time, we recover Eq.~(\ref{eq:2}) if $\delta_0< \Delta \lesssim u_\eta$, while if both $\Delta$ and  $\delta_0$ are  in the  inertial range we get
\begin{equation}
T_p(\Delta, \delta_0) \sim \frac{ T_0}{U_0^2} \left( \Delta^2 -  \delta_0^2 \right)\,, \label{eq:Tpredtutb}
\end{equation}
which is simply restating Lilly's argument (\ref{eq:lilly}) and Palmer et al.\cite{palmer2014} "real butterfly effect", which is nothing but  Lorenz's result quoted in the introduction. To recognize this, we can reason as follows. Let the initial uncertainty be as small as possible but larger than the Kolmogorov velocity, $\delta_0 \gtrsim u_\eta$, so that only the inertial range physics matter, which is the typical case in predictability problems in applications, then we can ask up to which time the uncertainty will be of the order of the velocity at the largest scales (i.e. $\Delta \lesssim U_0$)? To answer one can plug $\delta_0=u_\eta$ and $\Delta=U_0$ in Eq.~(\ref{eq:Tpredtutb}), to get $T_p\approx T_0 (1-u_\eta^2/U_0^2) =T_0(1-Re^{-1/2})\approx T_0$, where we used (\ref{eq:kolmo-u}). In other terms, the predictability time is finite and of the order of the large-scale eddy turnover time, as seen in Eq.~(\ref{eq:lilly}).

As already stated in the previous section, in principle one can increase $T_p$ by reducing the uncertainty to have $\delta_0$ in the dissipative range, but in that case the typically large Reynolds numbers and Eq.~(\ref{eq:delta0}) request the smallness of the initial uncertainty to be unrealistic (i.e. $\delta_0\lesssim \exp(-Re^{1/2})$. Anyway, one can still insist and try to pursue this path; in doing so, however, one can see that $\delta_0$ will reach values comparable with the thermal velocity, i.e., at which fluctuations induced by thermal noise will become relevant. In such a case, as discussed in Sec.~\ref{sec:noisySabra}, one has to consider the equation of fluctuating hydrodynamics. Clearly, one cannot have control over the thermal fluctuations, and so from the point of view of the predictability problem one has to consider systems with different noises \cite{loreto1996concept,boffetta2002predictability}, which amounts to having an uncertainty on the equations of motion. In Fig.~\ref{fig:fsle} we show with the full symbols the FSLE computed following two trajectories obtained from the noisy Sabra model (\ref{eq:Sabra-thermal}) with the same initial condition but different realizations of the noise (adding a perturbation smaller than the thermal velocity does not change the result). As one can see, at very small $\delta$,  $\lambda(\delta)$ is very high (it actually diverges for $\delta\to 0$) as it is dominated by noise, after it recovers the Lyapunov regime (as argued by Ruelle~\cite{ruelle1979microscopic}) and finally collapses onto the inertial $Re$-independent  $\delta^{-2}$ regime following the behavior of the FSLE computed in the deterministic shell model. This observation opens the issue of spontaneous stochasticity that will be discussed in Sec.~\ref{sec:spontaneous}, while in the next subsection we consider the fate of localized (in scale) perturbations.

%%%%%%%%%%%%%%%%%%%%%%%%%%%%%%%%%%%%%%%%%%%%%%%%%%%%%%%%%%%%%%%%%%%%%%%%%%%%%%%
\subsection{Growth of perturbations initially localized in the dissipative range\label{sec:preasymptotics}}
%%%%%%%%%%%%%%%%%%%%%%%%%%%%%%%%%%%%%%%%%%%%%%%%%%%%%%%%%%%%%%%%%%%%%%%%%%%%%%%

Recently, there has been some interest in the ``literal'' butterfly effect, i.e., the evolution of a small (infinitesimal) spatially localized perturbation. According to Ref.~\cite{shen2022three} (see also Refs.~\cite{pielke2024butterfly,Wolchover2011}), the perturbation, if initially in the dissipative range, should be re-adsorbed. We recall that the possibility of an initial decrease of the perturbation is not in contrast with the exponential growth~(\ref{eq:1}) which is expected at long times in a chaotic system. The exponential growth of the perturbation with the maximum LE requires the perturbation to be aligned with the maximal expanding direction (MED).  For a non-generic perturbation, especially in a system with many degrees of freedom, the initial evolution needed to align with the MED can be non-trivial.\cite{goldhirsch1987stability} During this pre-asymptotic regime, there are no constraints on the perturbation evolution, i.e., it can decrease. This poses the question concerning the duration of this initial transient time and its dependence on the initial spatial scale of the perturbation. Indeed, a closely related phenomenon has been recently shown in the context of a two-dimensional geophysical model, where an initially small and localized perturbation can lose energy at short times and then grow at later times.~\cite{SQG}

To investigate this behavior in shell models, we consider a small perturbation ${\bm u^\prime}$ which is initially localized in the dissipative range, e.g., at shell index $p$ such that $k_p>k_\eta$. The initial scale of the perturbation is $\ell_p \simeq 1/k_p$. The evolution of this infinitesimal perturbation is determined by the linearized equation
\begin{equation} 
\label{eq:model_linear}
\frac{d u^\prime_n}{dt}= i \mathcal{L}_n({\bm u^\prime},{\bm u}) - \nu k_n^2 u^\prime_n  \;,
\end{equation}
where $\mathcal{L}_n({\bm u^\prime},{\bm u}) = a k_{n+1}(u^{\prime *}_{n+1}u_{n+2}+u^*_{n+1}u^\prime_{n+2})  + b k_n(u^{\prime *}_{n-1}u_{n+1}+u^*_{n-1}u^\prime_{n+1}) - c k_{n-1}(u^\prime_{n-1}u_{n-2}+u_{n-1}u^\prime_{n-2})$.

The dynamics of the shell model (\ref{eq:Sabra}) is characterized by a strong temporal intermittency, which causes intense fluctuations in the perturbation evolution. As a consequence, the evolution of a single perturbation might not be representative of the typical behavior. In order to tackle this issue, we proceed as follows. First, we recast the evolution of the amplitude of the perturbation $\delta(t) = (\sum_n |u_n'(t)|^2)^{1/2}$ in terms of the finite-time Lyapunov exponents (FTLE):
\begin{equation} 
  \label{eq:gamma}
  \gamma_t = \frac{1}{t} \ln \left(\frac{\delta(t)}{\delta_0} \right) \ .\
\end{equation}
Then, we average the values of $\gamma_t$ over an ensemble of $N_e=10^3$ independent realizations of $({\bm u},{\bm u^\prime})$. The typical evolution of the perturbation is obtained as $\tilde{\delta}(t) = \delta_0 \exp(\langle \gamma_t \rangle t) = \exp(\langle \ln(\delta(t))\rangle)$. The advantage of using the exponential of the average of the logarithm instead of the average of the perturbation $\langle\delta(t)\rangle$ is that the latter is completely dominated by the fastest-growing replicas, while $\tilde{\delta}(t)$ allows us to properly take into account the contributions of all the replicas.

\begin{figure}[h!]
   \includegraphics[width=0.5\textwidth]{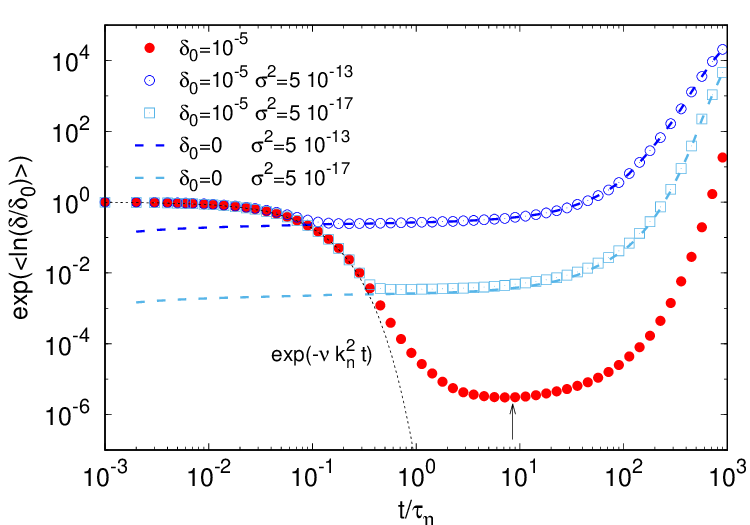}
 \caption{Time evolution of perturbation initially localized on a dissipative shell:  $\exp(\langle \ln(\delta(t))\rangle)$ vs time for the deterministic shell model (red filled circles), the stochastic shell model, evolving with different realizations of the noise, with $\sigma^2=5\cdot 10^{-13}$ (blue open circles) and with $\sigma^2=5\cdot 10^{-17}$ (light blue open squares) when the initial perturbation is localized on the shell $18$, i.e. at scales smaller than the Kolmogorov length with $\delta u_{18}(0)=10^{-5}$. The two dashed curves represent the evolution of two realizations ${\bm u_1}$ and ${\bm u_2}$ of the stochastic Sabra model, with $\sigma^2=5\cdot 10^{-13}$ (blue) and  $\sigma^2=5\cdot 10^{-17}$ (light blue), which starts from the same initial condition but evolve with different realization of the noise, for comparison with the other curves they have been normalized with the value $\delta_0=10^{-5}$ used in the other simulations. The results presented here are obtained with $N=23$ shells and $\nu=10^{-6}$ averaging over $10^3$ initial conditions/noise realizations. The perturbation size is computed summing over all the shells $\delta^2=\sum_{n} |u^\prime_n|^2$ with ${\bm u}^\prime = {\bm u_1} -  {\bm u_2}$. The arrow indicates the time $T_m$ at which the perturbation reaches its minimum in the deterministic case. \label{fig:localized}}
 \end{figure}

\begin{figure}[h!]
   \includegraphics[width=0.5\textwidth]{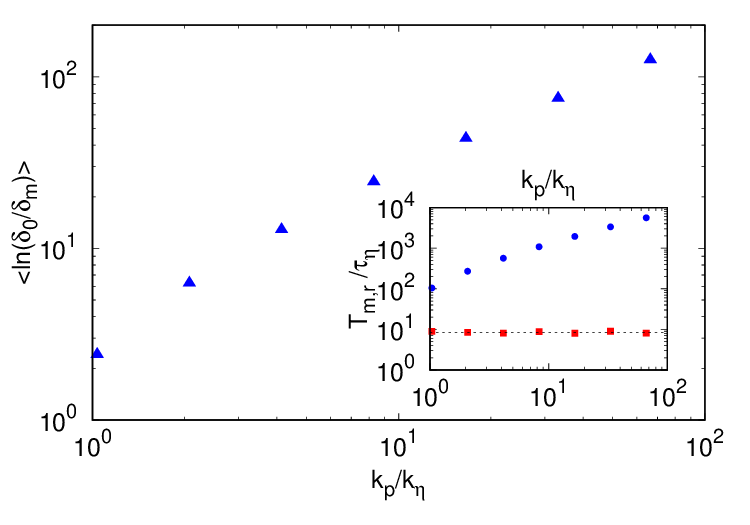}
 \caption{Average of the log of the ratio between initial and minimum
   values attained during the evolution of an infinitesimal
   \ perturbation initially localized in a dissipative shell $k_p$
   (blue triangles).  Inset: Time $T_m$ (red squares) at which the
   perturbation attains the minimum value $\delta_m$.  Time $T_r$
   (blue circles) at which the perturbation recovers its initial value
   $\delta(T_r)= \delta_0$.
 \label{fig:localized2}}
 \end{figure}

In Fig.~\ref{fig:localized} (with red circles) we show the typical evolution for a perturbation initially localized in the shell $p=18$, in the dissipative range. We observe that in the initial stage of the evolution, the amplitude of the perturbation decreases until it reaches a minimum $\delta_m$ at time $T_m$ (signaled by the small vertical arrow). We notice that the minimum value $\delta_m$ is very small when the wavenumber $k_p$ of the perturbation is larger than the Kolmogorov wavenumber, as reported in Fig.~\ref{fig:localized2}. At times $t> T_m$, the amplitude of the perturbation grows, and eventually it recovers its initial value $\delta_0$ at the "return time" $T_r$. The return time $T_r$ grows rapidly with the perturbed shell $k_p$, while $T_m$ remains almost constant $T_m \simeq (8.5 \pm 0.5 )\tau_\eta$. (as shown in the inset of Fig.~\ref{fig:localized2}).

The initial decrease of the perturbation is due to the viscous dissipation. This can be easily understood by a short-time Taylor expansion of Eq.~(\ref{eq:model_linear}) ${\bm u'}(t) = \bm {u'}(0) + t \left.(\partial {\bm u'}/\partial t)\right|_{t=0} + O(t^2) $, which gives $\gamma_t =  {\bm u'}(0)\cdot \left.(\partial {\bm u'}/\partial t)\right|_{t=0}/|{\bm u'}(0)|^2 + O(t)$. Averaging over independent random perturbations, we get $\langle \gamma_t \rangle =  - \nu k_p^2 + O(t)$, which shows that the typical initial growth rate is negative and that it depends on the viscosity and on the wavenumber of the perturbation $k_p$. The initial exponential decay of the perturbation due to the viscous force is clearly visible in Fig.~\ref{fig:localized} for $t < \tau_\eta$.

The results in Fig.~\ref{fig:localized2} show that the amplitude of the perturbation can decrease by several orders of magnitude during the initial phase of the evolution.  This implies that the perturbation can become of the order of the thermal noise; it is then natural to ask what happens. We address this question by considering the evolution of two realizations ${\bm u_1}$ and ${\bm u_2}$ of the stochastic Sabra model (\ref{eq:Sabra-thermal}) with different realizations of the thermal noise.  The perturbation field is defined as ${\bm u}^\prime = {\bm u_1} - {\bm u_2}$.

In this case, the initial decrease of the amplitude of the perturbation is expected to stop as soon as it reaches values where the uncertainty due to the different noise realizations starts to be relevant. After that, it should recover the same growth as that of the difference between two realizations starting from the same initial condition ($\delta_0=0$) and evolving with independent realizations of the noise. This is indeed confirmed by the results of numerical simulations of the stochastic Sabra model (\ref{eq:Sabra-thermal}) with two different intensity of the noise ($\sigma^2=5\cdot 10^{-13}$ and  $\sigma^2=5\cdot 10^{-17}$), which demonstrate that the initial decrease of the perturbation is bounded from below by the noise (see Fig.~\ref{fig:localized}).

Our results show that, as long as the validity of the deterministic equations can be assumed, a perturbation on scales smaller than the dissipative ones is always amplified exponentially over long times. The butterfly effect in the literal sense is therefore theoretically possible. However, the initial transient decay phase due to viscous forces can reduce the amplitude of the perturbation to the point where it is indistinguishable from thermal noise.

\section{Spontaneous stochasticity and predictability in turbulence\label{sec:spontaneous}}
In recent years, it has been recognized that the type of unpredictability discussed in the previous section and originating from Lorenz's ideas \cite{lorenz1969predictability} (quoted in the introduction) is deeply linked to a phenomenon that has been termed (Eulerian) \textit{spontaneous stochasticity}, \cite{mailybaev2016spontaneous,thalabard2020butterfly,bandak2024spontaneous} which is fundamentally tied to the non-uniqueness of solutions of the Euler equations \cite{daneri2021non}. In this section, we aim to show how the FSLE can be used to characterize this phenomenon. Actually, spontaneous stochasticity was first discovered in the Lagrangian context while considering the relative dispersion of particle tracers in turbulence,\cite{bernard1998slow,gawedzki2000phase,falkovich2001particles} i.e., the celebrated Richardson dispersion.\cite{richardson1926atmospheric} Since there is a tight link between the two forms of spontaneous stochasticity and the latter is somehow simpler to understand,  we will start by recalling the basic ideas of Lagrangian spontaneous stochasticity and showing how the FSLE can be used to characterize it.

\subsection{FSLE and Lagrangian spontaneous stochasticity\label{app:richardson}}

It is useful to start briefly recalling Richardson dispersion \cite{richardson1926atmospheric}, namely the evolution of the separation between two tracers transported by a turbulent flow, where the phenomenon of (Lagrangian) spontaneous stochasticity was first recognized \cite{bernard1998slow,gawedzki2000phase,falkovich2001particles}.

In 1926,  Richardson~\cite{richardson1926atmospheric}, while analyzing data of balloons in air turbulence, discovered that their separation grows as $\langle r^2(t) \rangle \propto \epsilon t^{3}$, which is the celebrated Richardson dispersion law.  The origin of this scaling law is rather clear if we use Kolmogorov theory. Indeed, the separation, $r$, between two particles, when $\eta<r<L_0$, is ruled by the velocity increments, $u_r$, over a distance $r$ and, after Kolmogorov, we know (neglecting intermittency) that $u_r\sim (\epsilon r)^{1/3}$ (see Sec.~\ref{sec:scales}). Then solving  $\dot{r}=u_r$ yields
\begin{equation}
r^{2}(t) \propto [2/3(r^{2/3}(0)+\epsilon^{1/3} t)]^3\,,
\label{eq:rich}
\end{equation}
which, for large $t$, reproduces Richardson scaling $r^2(t)\sim \epsilon t^3$.  For Eq.~(\ref{eq:rich}) to apply, however, one has to require $r(0)>\eta$ because, for $r(0)<\eta$, the velocity field would be smoothed by viscosity, $u_r \propto r$, leading to chaos to come into play with an exponential growth of the separation $r(t) \approx r(0) \exp(\lambda t)$ (lasting until $r(t)<\eta$), with $\lambda$ (at least dimensionally) given by Eq.~(\ref{eq:ruelle}).  However, increasing the Reynolds number (i.e., in the limit $\nu \to 0$) from Eq.~(\ref{eq:kolmo-l}), we know that $\eta\to 0$, therefore it is natural to ask whether we can still use Eq.~(\ref{eq:rich}) and perform the limit $r(0)\to 0$.  Setting $r(0)=0$ in Eq.~(\ref{eq:rich}) and posing $\nu=0$, we indeed still find Richardson result. However, the equation $\dot{r}=u_r$ is also solved by $r(t)\equiv 0$. This is because having $u_r \propto r^{1/3}$ means that the velocity is rough, i.e., not Lipschitz continuous, and consequently the Cauchy theorem for the uniqueness of solutions does not hold (see also Ref.~\cite{vanpoucke2021assigning} for a nice example of the same phenomenon in a mechanical context).

The connection between non-uniqueness of solutions and Richardson dispersion was first recognized in Ref.~\cite{bernard1998slow} (see also Refs.~\cite{gawedzki2000phase,falkovich2001particles}) in the context of the Kraichnan model~\cite{kraichnan1968small} for passive scalar transport, which allows for rigorous treatment.  Mathematically, the issue boils down to the order in which the limits $r(0)\to 0$ and $\nu\to 0$ are performed. For better framing the issue, it is useful to consider the presence of noise; physically, this is justified by the fact that (small) tracer particles experience Brownian motion due to thermal agitation. Including noise, the particle separation is ruled by a Langevin equation that, for the sake of simplicity, we can write as $\dot{r}=u_r + \sqrt{4\kappa} \phi$ (the factor $4$ in the square root accounts for the fact that we are considering the separation between two particles), where the noise $\phi$ can be assumed as usual a Gaussian zero-mean and unit variance, time uncorrelated process. The presence of noise has the nice mathematical property of regularizing the velocity field for $r\to 0$: given the realization of the noise, the solution to the above stochastic equation is unique (see, e.g., Ref.\cite{flandoli2013topics} for a mathematical discussion).  Essentially, the presence of noise acts as a point splitting of the particle pair \cite{frisch1999lagrangian} and thus a regularization of the singularity at $r=0$, when $\eta=0$. Now with the noise we can safely take $r(0)=0$, and, for $\nu,\kappa >0$, by considering many realizations of the noise (i.e. many pairs starting in the same point) we have a distribution of solutions that at time $t$ are described by a density $\rho_{\kappa,Re}(r,t|0)$ labeled by the values of diffusivity and the Reynolds number (i.e. the inverse of the viscosity). Now the question is how $\rho_{\kappa,Re}(r,t|0)$ behaves when $Re\to \infty$ ($\nu\to 0$) and $\kappa\to 0$, and, in particular, how it depends on the order in which the two limits are taken.

It is natural to expect that if $Re \!\to\! \infty$ after $\kappa \!\to\! 0$ then $\rho_{\kappa,Re}(r,t|0)\! \to \!\delta(r(t))$, i.e., the limit recovers the deterministic dynamics with uniqueness of the solution guaranteed by the viscous cutoff. While it can be proved~\cite{gawedzki2000phase,falkovich2001particles} that if $\kappa \!\to\! 0$ after $Re \!\to\! \infty$ then $\rho_{\kappa,Re}(r,t|0)\! \to\! \rho^*(r,t|0)$ with $\rho^*(r,t|0)$ a well-defined distribution with variance growing according to Richardson law, i.e. $\langle r^{2}(t)\rangle= \int dr r^2 \rho^*(r,t|0)\propto t^3$.  In other terms, the two limits
\begin{equation}
\lim_{\kappa \to 0} \lim_{Re \to \infty} \neq \lim_{Re
  \to \infty} \lim_{\kappa \to 0}\, \label{eq:limits}
\end{equation}
do not commute, and with the ordering on the l.h.s., the solutions are not unique and remain stochastic. This is, without any claim of mathematical rigor (which can be found in the original works\cite{bernard1998slow,gawedzki2000phase}), the essence of the (Lagrangian) spontaneous stochasticity. It is worth stressing that a byproduct of the previous observations is that particle separation is explosive: Eq.~(\ref{eq:rich}) tells us that an initially coincident (physically speaking $r(0)\gtrsim \eta$) pair of particles separates to distances of order of the integral scale $L_0$ in finite time, which should immediately suggest a link with Eq.~(\ref{eq:Tpredtutb}).

Below, we connect Lagrangian spontaneous stochasticity with the FSLE.  Let us first introduce a new scale
\begin{equation}
\ell_\kappa=\sqrt{\kappa \tau_\eta} = \eta \sqrt{\kappa/(u_\eta\eta)}=
\eta Pe^{-1/2}Re^{1/2} \ ,\
\label{eq:ellek}
\end{equation}
where we used (\ref{eq:kolmo-l}) and (\ref{eq:kolmo-u}), while $Pe=U_0L_0/\kappa$ is the Peclet number. For $r<\ell_\kappa$, the particle separation is dominated by noise, above it by the advection due to the velocity field.  Let us consider now the behavior of the FSLE. Let us call $\mu^{Re,\kappa}(r)$ the Lagrangian FSLE for the separation of two tracers; we should expect
\begin{subequations}\label{eq:lag-fsle}
\begin{empheq}[left={\mu^{Re,\kappa}(r)\simeq\empheqlbrace}]{align}
&4\kappa r^{-2}
&& \text{if } r\lesssim \ell_\kappa
\label{eq:diffFSLE}\\
&\lambda
&& \text{if } r\lesssim \eta
\label{eq:lyapFSLE}\\
&\epsilon^{1/3}r^{-2/3}
&& \text{if } \eta<r<L_0
\label{eq:richFSLE}
\end{empheq}
\end{subequations}
where  (\ref{eq:richFSLE}) is obtained by dimensional arguments. In order to establish a  relationship between $\ell_\kappa$ and $\eta$, we first  notice that from Eq.~(\ref{eq:ellek})
  \begin{equation}
  \frac{\ell_\kappa}{\eta}= Pe^{-1/2} Re^{1/2}= \frac{1}{Sc^{1/2}}=\left(\frac{\kappa}{\nu}\right)^{1/2}\,,
  \end{equation}
where $Sc=\nu/\kappa$ is the Schmidt number.  We can now translate the two limit procedures (\ref{eq:limits}). Performing $\lim_{Re\to \infty} \lim_{\kappa\to 0}$ corresponds to $Sc\to \infty$ i.e. $\kappa \to 0$ faster than $\nu\to 0$. In such case $\ell_\kappa \to 0$ faster than $\eta \to 0$, which means that $\lim_{Re\to \infty}\lim_{\kappa \to 0} \mu^{Re,\kappa}(r) =\lambda$ so that (even if $\lambda \to \infty$ for $Re\to \infty$) the limit is deterministic and we have uniqueness of the trajectories. The opposite happens with when $\lim_{\kappa\to 0} \lim_{Re \to \infty}$ in this case asymptotically $\ell_\kappa > \eta$ and thus
\begin{equation}
\lim_{\kappa\to 0}\lim_{Re \to \infty} \mu^{Re,\kappa}(r)
=\frac{\epsilon^{1/3}}{r^{2/3}} \,,\label{eq:fsle-rich-sp}
\end{equation}
for the latter to hold true is enough that $Pe\to \infty$ slower than $Re \to \infty$, which corresponds to $Sc\to 0$.  Nevertheless, we can expect that also for finite $Sc$ at sufficiently large scales the behavior is independent of noise and of $Re$. We also note that if asymptotically Eq.~(\ref{eq:fsle-rich-sp}) holds, then using Eq~(\ref{eq:fsle-tpred}) we clearly see that even if the initial separation is initially zero, the time it takes for the separation to reach the scale $R$ is
\begin{equation}
  \mathcal{T}(R,0) \propto
  \frac{R^{2/3}}{\epsilon^{1/3}}=\tau_R\,, \label{eq:explosive}
\end{equation}
i.e., it is finite and proportional to the eddy turnover time of the scale $R$ (see Eq.~(\ref{eq:tempivel}).

To illustrate the above considerations, we show below the FSLE for particle separation, computed with the Kraichnan model~\cite{kraichnan1968small} in which the velocity field is taken as a Gaussian zero-mean and time-uncorrelated field, with spatial correlations that mimic the scaling laws known for turbulent velocities (see Ref.~\cite{falkovich2001particles} for a thorough review on the subject). In particular, we consider a (small-scale) regularized version of the original Kraichnan model, i.e., we write the tensor controlling the statistics of the velocity differences as
\begin{eqnarray}
  d_{ij}(\bm r)\!\! &=& \!\!D_1(r^2\! +\! \eta^2)^{\frac{\xi}{2}} \left[ \left( d\! -\! 1\!+\! \xi \frac{r^2}{r^2\! +\! \eta^2} \right) \delta_{ij} - \xi \frac{r_i r_j}{r^2 \!+\! \eta^2} \right]\nonumber
 \\ &-&\!\!D_1(d-1)\eta^{\xi} \delta_{ij}\,, \label{eq:varkraich}
\end{eqnarray}
where $d$ denotes the space dimensionality, $\xi$ the roughness exponent, $\eta$ the viscous cut-off and $D_1$ is a constant with dimension $(length)^{2-\xi}/(time)$, measuring the velocity intensity.  The choice (\ref{eq:varkraich}) ensures incompressibility ($\sum_i \partial_i d_{ij}(\bm r)=0$) and nicely interpolates between the smooth (with $d_{ij}(\bm r)=D_1 \xi \eta^{\xi-2}/2\, r^2\,[(d+1)\delta_{ij}-2r_ir_j/r^2]$) and rough  (with $d_{ij}(\bm r)=D_1 r^\xi[(d-1+\xi)\delta_{ij}-\xi r_ir_j/r^2]$) Kraichnan model in the limits $r\ll \eta$ and $r\gg \eta$, respectively.  Therefore, we expect~\cite{falkovich2001particles} exponential growth of $r(t)$ for $r(t)\lesssim \eta$ with Lyapunov exponent $\lambda=D_1\xi [d(d-1)/2] \eta^{\xi-2}$ and Richardson-like growth, $r^2(t) \propto t^{2/(2-\xi)}$ for $r(t)\gtrsim \eta$. Notice that the choice $\xi=4/3$ leads to Richardson scaling $t^{3}$ and will be used in the numerical computations.

\begin{figure}[t!]
\includegraphics[width=0.5\textwidth]{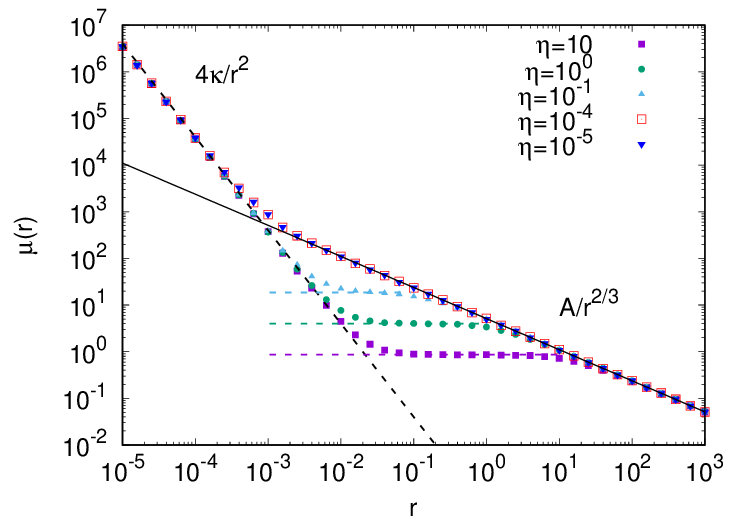}
 \caption{FSLE $\mu(r)$ for the particle separation for different values of $\eta$ as in the legend and $\kappa=10^{-4}$. With such value of $\kappa$ the diffusion dominates the inertial range physics for $r<r_\kappa=10^{-3}$, indeed when $\eta$ is above such value we still see the Lyapunov regime (dashed horizontal lines displays the analytical value of $\lambda$), while when $\eta$ is below such value only Richardson ($\mu(r) \sim r^{-2/3}$) and diffusive ($\mu(r) \approx 4\kappa r^{-2}$) behaviors are visible and $\mu$ becomes independent of $\eta$. Clearly, if we then perform the limit $\kappa\to 0$ we have $\mu(r)\to A r^{-2/3}$, with $A=D_1d(d-1)(2-\xi)\ln\beta/(\beta^{2-\xi}-1)$ as can be obtained analytically solving for the mean exit time, where $\beta$ is the scale separation between consecutive separations considered in the evaluation of the FSLE, i.e. $\beta=R_n/R_{n-1}$.  Here the FSLE was computed using the procedure described in Eq.~(\ref{eq:app10}) of Appendix~\ref{app:fsle} with  $\xi=4/3$ and $D_1=1$ for $d=3$. \label{fig:richardson}}
 \end{figure}

We can now consider the SDE for the particle separation, which reads:
\begin{equation}
dr_i = \mathbb{L}_{ij} dW_j + \sqrt{4\kappa} dB_i \qquad i,j=1,\ldots,d
\label{eq:sde}
\end{equation}  
where $(\mathbb{L}\mathbb{L}^T)_{ij}=2d_{ij}$ with $\mathbb{L}$ obtained by performing the Cholesky decomposition of $d_{ij}$\cite{frisch1999lagrangian}, $dW_i$ and $dB_i$ are independent Wiener processes.  The presence of the diffusive term on the r.h.s. with diffusivity $\kappa$ introduces a new scale $r_\kappa \approx \eta [\kappa/(\eta^{\xi}D_1)]^{1/2}$ when $r_\kappa <\eta$ and $r_\kappa \propto (\kappa/D_1)^{1/\xi}$ when $r_\kappa>\eta$. Clearly, $r_\kappa$ plays the same role as $\ell_\kappa$ in Eq.~(\ref{eq:ellek}).  Computing the FSLE, $\mu(r)$, following the limit procedure $\lim_{\kappa \to 0} \lim_{\eta\to 0}$ corresponds to the limit leading to Lagrangian spontaneous stochasticity. Such a limit is illustrated in Fig.~\ref{fig:richardson}, which displays the FSLE computed with fixed $\kappa$ at decreasing $\eta$. The figure clearly shows that when $r_\kappa/\eta>1$ the plateau corresponding to $\mu(r)=\lambda$ disappears and only the diffusive (\ref{eq:diffFSLE}) and Richardson (\ref{eq:richFSLE}) regimes survive. Then, clearly performing the limit $\kappa\to 0$, only the Richardson regime survives, as it corresponds to moving to zero the point $r_\kappa=(\kappa/D_1)^{1/\xi}$ where the behaviors (\ref{eq:diffFSLE}) and (\ref{eq:richFSLE}) intersect, thus demonstrating the result (\ref{eq:fsle-rich-sp}).

\subsection{FSLE and Eulerian Spontaneous Stochasticity}

In the previous section, we have seen how the explosive separation between particle tracers in turbulent flows is a manifestation of the non-uniqueness of particle trajectories in the limit of vanishing viscosity and that this is well described by the limiting behavior of the finite-size Lyapunov exponent. For the velocity field, something very similar happens and, again, can be characterized via the FSLE as shown in the following in the context of the shell model, even though the argument extends, unchanged, to the Navier-Stokes equation.

Let us consider replicas of the Sabra model with the same initial conditions and different realizations of the noise, this is like considering an initial perturbation of the order of the thermal velocity, $\delta_0 \simeq u_{th}$ (where, with reference to Eq.~(\ref{eq:Sabra-thermal}) $u_{th}\approx \sigma$), which is well inside the dissipative range, namely in the plateau region of Fig.~\ref{fig:moments}. As seen in Fig.~\ref{fig:fsle}, indeed there is a $\delta^{*} \gtrsim u_{th}$, independent of $Re$, above which the FSLE behave as in the deterministic Sabra model, with $\lambda(\delta) \approx \lambda$, and then, for $\delta>u_\eta$ behaves as $\lambda(\delta)\approx \epsilon \delta^{-2}$. We can then ask the time, $\mathcal{T}$, it takes the perturbation induced by thermal noise to reach a value $\Delta$ in the inertial range, i.e., for the initially small perturbation to perform an inverse cascade up to affecting scales of the order $\ell\sim \Delta^{3}/\epsilon$. There are two contributions\footnote{As discussed in Sec.~\ref{sec:preasymptotics}, in principle, a third contribution due to the possible transient decrease of the perturbation must be considered, but this is just adding a constant contribution when noise is present (Cfr. Fig.~\ref{fig:localized}.}  to the duration of this time, $\mathcal{T}=\mathcal{T}_1+\mathcal{T}_2$. Clearly, at the beginning, there will be a regime, lasting $\mathcal{T}_1$, controlled by the sensitive dependence on initial conditions due to chaos, as originally proposed by Ruelle.\cite{ruelle1979microscopic}. The exponential growth lasts until the velocity difference starts to be of the order of the Kolmogorov velocity, $\delta(\mathcal{T}_1)\approx u_\eta$, after which the inertial range physics comes into play. The duration of this regime is given by Eq.~(\ref{eq:fsle-tpred}) with the integral extrema $\delta_0=u_{th}$ and $\Delta=u_\eta$, since in this range $\lambda(\delta)=\lambda$ the result is given by Eq.~(\ref{eq:2}), i.e.:
\begin{equation}
\mathcal{T}_1=\left(\frac{\nu}{\epsilon}\right)^{1/2} \ln\left(\frac{u_{\eta}}{u_{th}}\right)=T_0 Re^{-1/2} \ln\left(\frac{u_{\eta}}{u_{th}}\right)\,,
  \label{eq:T1}
\end{equation}  
where we used Eqs.~(\ref{eq:ruelle}) and (\ref{eq:kolmo-t}). Once $\delta\approx u_\eta$ we can use again Eq.~(\ref{eq:fsle-tpred}) now posing $\delta_0=u_\eta$ and $\lambda(\delta)=\epsilon/\delta^{2}$, to get the second contribution $\mathcal{T}_2$, which is  Eq.~(\ref{eq:Tpredtutb}) that we can rewrite here as
\begin{equation}
  \mathcal{T}_2 \!\approx\! \frac{1}{\epsilon} \left( \Delta^2\! -\!  u_\eta^2 \right)=
  \frac{\Delta^2}{\epsilon}\!-\!\left(\frac{\nu}{\epsilon}\right)^{1/2}
  \!=\!\frac{\Delta^2}{\epsilon}\!-\!T_0 Re^{-1/2}
  \label{eq:T2} \ ,\
\end{equation}  
where in the second and third equalities we used Eq.~(\ref{eq:kolmo-u}). Using  Eqs.~(\ref{eq:T1}) and (\ref{eq:T2}) we see that the time for the thermal fluctuations, which start in the dissipative range well below the Kolmogorov scale, to reach the scale $\ell=\Delta^{3}/\epsilon$ in the inertial range is
\begin{eqnarray}
  \mathcal{T} &=& T_0 \left(\frac{\ell}{L_0}\right)^{2/3} + T_0 Re^{-1/2}\left[\ln\left(\frac{u_{\eta}}{u_{th}}\right)-1\right] \nonumber\\
&=&  \tau_\ell+ T_0 Re^{-1/2}\left[\ln\left(\frac{u_{\eta}}{u_{th}}\right)-1\right]
  \underbrace{\to}_{Re\to \infty} \tau_\ell\,,
\label{eq:Tf}
\end{eqnarray}
i.e., simply the eddy turnover time of that scale (where we used Eq.~(\ref{eq:tempivel})). And, for $\ell=L_0$ we have that $\mathcal{T}\approx T_0$, again confirming Lorenz's and Lilly's arguments in an even stronger form: the uncertainty induced by thermal fluctuations is enough to propagate to larger scales in a finite time bounded by the eddy turnover time of the scale of interest.  Notice also that Eq.~(\ref{eq:Tf}) is basically equivalent to the result (\ref{eq:explosive}) we found for Richardson dispersion, again demonstrating the deep link between the two forms of spontaneous stochasticity. Heuristically, rephrasing a nice argument from Ref.~\cite{Dubrulle}, this link can be understood as follows. The fact that $\lambda(\delta) \propto \epsilon/\delta^{-2}$ says that $\delta \sim (\epsilon t)^{1/2}$ (which is another way to read Eq.~(\ref{eq:T2}). Now the velocity difference $\delta$ can be seen as the derivative of a separation $\delta \sim \dot{r}\sim (\epsilon t)^{1/2}$ meaning that $r\sim \epsilon^{1/2} t^{3/2}$ and thus $t \sim r^{2/3} \epsilon^{1/3}$, which implies $\mu(r)=\epsilon^{1/3}/r^{2/3}$, that is Eq.~(\ref{eq:fsle-rich-sp}).

For Eq.~(\ref{eq:Tf}) to apply it is enough to require that the noise goes to zero slowly enough such that $Re^{-1/2}\ln({u_{\eta}}/{u_{th}}) \to 0$, which amounts to require $u_{th}$ going to 0 slower than $\exp(-Re^{1/2})$.  In terms of the FSLE shown in Fig.~\ref{fig:fsle} this means that denoting with $\delta^{*}\gtrsim u_{th}$ where the the FSLE computed with noise matches the Lyapunov exponent $\lambda(\delta)\approx \lambda$, the $Re\to \infty$ and $u_{th}\to 0$ (zero noise) limits should be done in such a way that the interval $[\delta^{*},u_{\eta}]$ shrinks to zero faster than $\delta^{*}\to 0$. In this limit $\lambda(\delta) \to \epsilon \delta^{-2}$ for any $\delta$; again, this is the same idea we used in the previous section for Richardson dispersion (C.f.r. Fig.~\ref{fig:richardson}).  When performing the limit $Re\to \infty$ and vanishing noise as discussed above, the predictability time at scale $\ell$ or, equivalently, for tolerance on the velocity field of order $\Delta\approx (\epsilon \ell)^{1/3}$ is finite and basically given by the eddy turnover time $\tau_\ell=T_0(\ell/L_0)^{2/3}$ of that scale or associated to that velocity tolerance, which is the same via Eq.~(\ref{eq:tempivel}). This is the manifestation of the fact that the solution of the system for zero viscosity is not unique and thus of  the (Eulerian) spontaneous stochasticity, which was nicely demonstrated in direct numerical simulations of flows generated by the Kelvin–Helmholtz instability of a singular shear layer.\cite{thalabard2020butterfly}

We acknowledge that the arguments that lead us from Eq.~(\ref{eq:T1}) to (\ref{eq:Tf}) are basically the same as those that can be found in \cite{bandak2024spontaneous} but, here, rephrased with the FSLE.  In Ref.~\cite{bandak2024spontaneous} it is shown that, independently of the Reynolds number, given the presence of thermal noise, the distribution of solutions obtained with different realizations of the noise is ``unique'', as for the particle separation discussed in the previous section~\cite{flandoli2024remarks}.  In particular, in Ref.~\cite{bandak2024spontaneous} two cases were considered, one in which the initial condition of the shell velocities is the Kolmogorov solution $u_n \propto k_n^{-1/3}$ and one in which it is an arbitrary configuration taken from the stationary dynamics. It was shown that in both cases the distribution of solutions is independent of the noise and of $Re$. Moreover, it was shown that the variance of such distribution asymptotically grows in time. In the case of the Kolmogorov initial condition it grows linearly while in the other case with a slightly different exponent, and this was linked to the fact that, due to intermittency, the latter was a state deviating from Kolmogorov scaling, i.e. $u_n \propto k_n^{-h}$ with $h\neq 1/3$ and thus characterized by a time $\tau_n \propto k_n^{-(1-h)}$. Using Eq.~(\ref{eq:T2}) or equivalently Eq~(\ref{eq:Tpredtutb}) we can see that asymptotically $\Delta^2 \propto \epsilon t$, this has precisely the same meaning of the growth of the variance of the distribution of solutions and directly derive from Eq.~(\ref{eq:fsle-tpred}) and the $\lambda(\delta)=\epsilon \delta^{-2}$. The only difference between this derivation based on FSLE and the one discussed in Ref.~ \cite{bandak2024spontaneous} is that the FSLE is an average quantity, namely it averages the way the spread of the solutions over all possible initial conditions, and this average leads to the linear growth because it is not affected by intermittency corrections, as shown in Appendix~\ref{app:fsle}.

%%%%%%%%%%%%%%%%%%%%%%%%%%%%%%%%%%%%%%%%%%%%%%%%%%%%%%%%%%%%%%%%%%%%%%%%%%%%%%%
\section{Discussions and conclusions\label{sec:discussions}}
%%%%%%%%%%%%%%%%%%%%%%%%%%%%%%%%%%%%%%%%%%%%%%%%%%%%%%%%%%%%%%%%%%%%%%%%%%%%%%%
Over half a century ago, Edward Lorenz introduced the evocative metaphor of the ``butterfly effect'', namely the idea that a butterfly flapping its wings in Brazil could ultimately trigger a tornado in Texas. Today, this metaphor is deeply embedded in popular culture, where it is commonly explained mathematically through the existence of a positive maximal Lyapunov exponent ($\lambda>0$). However, as understood by Lorenz himself and stressed in this work, this standard dynamical systems perspective is fundamentally incomplete when applied to fully developed, multiscale turbulent flows.

The limitation of the traditional Lyapunov exponent lies in its definition, which restricts its relevance to infinitesimal perturbations. In high-Reynolds-number turbulence, this chaotic regime is confined to sub-Kolmogorov scales, where the characteristic velocity is physically negligible. For any macroscopic forecasting attempt, the tolerances and initial uncertainties are inevitably much larger than the Kolmogorov scale. Consequently, predicting such systems requires understanding how finite-size perturbations grow. Historically, this was pioneered by Lilly,\cite{lilly1972numerical} Lorenz\cite{lorenz1969predictability,lorenz1996predictability} and others through phenomenological arguments of an ``error cascade'' traveling upward through the inertial range. More recently, it was recognized that this classical picture is tightly linked to the phenomenon of spontaneous stochasticity,\cite{palmer2014,mailybaev2016spontaneous,thalabard2020butterfly,bandak2024spontaneous} which ultimately boils down to the non-uniqueness of the solutions in the limit of infinite Reynolds number and causes even infinitesimally small perturbations induced by thermal noise to reach the large scales in finite time.\cite{bandak2024spontaneous}

In this work, we have revisited the classical approaches to predictability in turbulence and their modern developments using the Finite Size Lyapunov Exponent\cite{aurell1996growth,aurell1997predictability,cencini2013finite} as a bridge between these paradigms, and used shell models as a laboratory for the multiscale physics of real turbulent flows. In the limit of infinitesimal perturbations, $\lambda(\delta)$ recovers the standard Lyapunov exponent. For finite perturbations in the inertial range, however, it reveals a scale-dependent growth rate ($\lambda(\delta)\approx \epsilon\delta^{-2}$) that is independent of the Reynolds number. Notably, the FSLE can also be used to characterize the evolution of replicas with different realizations of the noise, thus providing a way to characterize the subtle aspects emerging when considering the limits of fully developed turbulence, vanishing initial perturbations and noise strength, whose non-commutativity is not a mere mathematical curiosity but the definitive signature of \textit{spontaneous stochasticity}. Indeed, we have shown that, in both Eulerian (with the shell model) and Lagrangian (with the Kraichnan model) spontaneous stochasticity --- two facets of the emergence of non-deterministic behaviors in turbulence ---, the FSLE attains a precise limiting functional form characterizing the intrinsic unpredictability of turbulent flows.

By organizing these interconnected ideas—from the Kraichnan model to the Sabra shell model—under the unified umbrella of the FSLE, we have shown how this modern paradigm bridges the gap between Edward Lorenz's intuitions and the rigorous mathematics of modern statistical physics. The "real" butterfly effect is not just about sensitivity to initial conditions—it is the realization that at high Reynolds numbers, nature is fundamentally and spontaneously stochastic.

Furthermore, stimulated by recent interest in the so-called ``literal butterfly'' effect\cite{shen2022three,pielke2024butterfly,Wolchover2011} we have also considered infinitesimal perturbations localized within the dissipative range, and showed that they undergo a transient, viscous-driven decay before aligning with the expanding directions of the chaotic attractor. Yet, in the real world, this decay is never absolute. When thermal fluctuations are included in the description, via fluctuating hydrodynamics, they act as an insurmountable floor, rescuing perturbations from vanishing and feeding them back into the upward macroscopic cascade.

\section*{Data availability statement}
The data that supports the findings of this study are available from the corresponding author upon reasonable request.

\begin{acknowledgments}
We wish to acknowledge useful discussions with  Alexei Mailybaev. V. J. V. acknowledges the support by the Italian Ministry of University and Research (MUR) - Fondo Italiano per la Scienza (FIS2) - 2023 Call, project DeepFL, CUP: E53C24003760001.
\end{acknowledgments}

\appendix

\section{Lyapunov exponent and FSLE in turbulence}
\label{app:fsle}
In this appendix, we provide some details on the computation of the Finite Size Lyapunov exponent and also consider the effects of intermittency using the multifractal formalism~\cite{frisch95}. We start by reconsidering Ruelle's prediction (\ref{eq:ruelle}) for the Lyapunov exponent to show that, while dimensionally correct, intermittency affects its estimation; then we extend this approach to the FSLE to show that the $\delta^{-2}$ behavior is insensitive to intermittency correction and finally provide some ideas on how the FSLE can be computed in practice.

As suggested by Ruelle \cite{ruelle1979microscopic}, the Lyapunov exponent in turbulence is expected to be proportional to the inverse of the shortest dynamical time, i.e., the Kolmogorov time $\tau_{\eta}$.  Since this is a small-scale quantity, we expect that it is affected by intermittency and that the dimensional prediction (\ref{eq:ruelle}) has to be corrected. These corrections can be easily understood by using the multifractal model of turbulence \cite{frisch95}, which assumes that velocity fluctuation $u_{\ell}$ at the scale $\ell$ in the inertial range has the scaling exponent $h$, i.e.  $u_{\ell} \simeq U_0 (\ell/L_0)^h$, with probability given by $P_{\ell}(h) \sim (\ell/L_0)^{3-D(h)}$, where $D(h)$ is an universal function which can be measured in experiments or numerical simulations.  By using the definition of the Kolmogorov scale $u_{\eta} \eta/\nu=1$ and that dimensionally $\tau_{\eta}=\eta/u_{\eta}$, we can write
\begin{equation}
\lambda \!\simeq\! \int\! dh \frac{ P_{\eta}(h)}{\tau_{\eta(h)}} \!\simeq\!
\frac{1}{T_0} \int  dh  \left(\frac{\eta}{L_0}\right)^{h+2-D(h)}\!\!\!\!\!
\simeq \!\frac{1}{T_0} Re^{\alpha}\,.
\label{eq:app1}
\end{equation}
Using that $Re=U_0 L_0/\nu \simeq (L_0/\eta)^{1+h}$ yields
\begin{equation}
\lambda \simeq \frac{1}{T_0} \int Re^{(D(h)-h-2)/(1+h)} dh 
\simeq \frac{1}{T_0} Re^{\alpha}\,,
\label{eq:app2}
\end{equation}
where the integral, in the limit of large $Re$, can be estimated by the saddle point to give
\begin{equation}
\alpha=\max_h \left\{\frac{D(h)-2-h}{1+h}\right\}\,.
\label{eq:app3}
\end{equation}
Using the $D(h)$ provided by the She-Leveque model of intermittency~\cite{sheleveque}, one obtains $\alpha \simeq 0.47$, close but different from the dimensional estimation $\alpha=1/2$. In the shell model, this value was confirmed~\cite{crisanti1993intermittency}. In 3D turbulence, recent results found a different value~\cite{boffetta2017chaos,berera2018chaotic,banerjee2026intermittent}, the origin of which is still debated and may be linked to effects coming from sweeping of the small scales from the large ones~\cite{ge2023production}, an effect that is absent in the shell model.

The above estimation can be extended to the computation of the Finite Size Lyapunov Exponent when the perturbation is within the inertial range of scales \cite{aurell1996growth}.  In this case, we assume that $\lambda(\delta)$ is proportional to the inverse of the eddy turnover time at the scale $\ell$, which corresponds to a velocity fluctuation $u_\ell \simeq \delta$.  Since $\tau_{\ell} \simeq T_0 (\ell/L_0)^{h-1} \simeq T_0 (u_{\ell}/U_0)^{1-1/h}$ we can write
\begin{equation}
\lambda(\delta) \simeq \frac{1}{T_0} \int dh\left(\frac{\delta}{U_0}\right)^{1-1/h} P(h) \,,
\label{eq:app4}
\end{equation}
which, plugging  the expression for the probability, yields
\begin{equation}
\lambda(\delta) 
\simeq \frac{1}{T_0} \int dh \left(\frac{\delta}{U_0}\right)^{(2+h-D(h))/h} \!\!\!\!
\simeq \frac{1}{T_0} \left(\frac{\delta}{U_0}\right)^{\beta}\,,
\label{eq:app5}
\end{equation}
where again we can use a saddle point estimation (valid when $(\delta/U_0) \ll 1$) for the exponent
\begin{equation}
\beta = \min_h \left\{\frac{2 + h - D(h)}{h}\right\}\,.
\label{eq:app6}
\end{equation}
By using the standard inequality for the multifractal model \cite{frisch95} (a consequence of the Kolmogorov $4/5$ law), $D(h) \le 3h+2$ (with the equality realized for the exponent associated to the third-order velocity structure function), we can write
\begin{equation}
\frac{2 + h - D(h)}{h} \ge -2\,,
\label{eq:app7}
\end{equation}
and therefore from (\ref{eq:app6}) we have $\beta=-2$, i.e., the dimensional prediction even in the presence of intermittency. This result can be physically understood by observing that in (\ref{eq:fsle}) $\lambda(\delta)$ depends on the energy dissipation rate $\varepsilon$ to the power one, which is not affected by intermittency corrections.

The Finite Size Lyapunov Exponent is defined as a generalization of the Lyapunov exponent starting from the expression for the predictability time (\ref{eq:2}). From the predictability time $T(\delta_0,\Delta)$ for a perturbation to grow from $\delta_0$ to $\Delta$, we can define the FSLE as
\begin{equation}
\lambda(\delta_0,\Delta)\equiv \left\langle {1 \over T(\delta_0,\Delta)} 
\ln \left({\Delta \over \delta_0} \right) \right\rangle
\label{eq:app8}
\end{equation}
averaged along a very long reference trajectory. This is a generalization in the sense that when both $\delta_0$ and $\Delta$ are infinitesimal, it recovers the Lyapunov exponent.

When both $\delta_0$ and $\Delta$ are finite, Eq.~(\ref{eq:app8}) averages the contributions from all the scales between $\delta_0$ and $\Delta$, which typically have different growth rates. To measure the growth rate at a given scale $\delta$, one must keep both $\delta_0$ and $\Delta>\delta_0$ of the order of $\delta$. This can be obtained by slightly different procedures, which we discuss here for the case of the shell model. We remark here a delicate point, i.e., that the two trajectories which are used to define the error $\delta$ have to be on the attractor of the system. While this condition is easily satisfied for the standard Lyapunov exponent (e.g., by working in the tangent space of infinitesimal separations), in the case of finite separation, this request needs to be carefully implemented.

We first discretize the scales of the perturbation in a discrete series of thresholds $\delta_n=\delta_0 r^n$ with a rate $r>1$.  The rate cannot be large to separate the contribution of the different scales but, of course, should be larger than $1$. A reasonable choice is, e.g., $r=2$.

Starting from a statistically stationary configuration $\bm{u}=\{u_n\}$, we add a very small perturbation to generate a new configuration $\bm{u}'$. We then integrate both configurations until their separation $\delta=\left[\sum_n |u'_n-u_n|^2 \right]^{1/2}$ reaches the first threshold $\delta_0$ (and both trajectories are on the attractor).  We then compute the {\it doubling times}, i.e., the times $T(\delta_n)$ it takes the separation to grow from the threshold $\delta_n$ to the next $\delta_{n+1}$. When the separation grows beyond the last threshold, we generate a new perturbed configuration $\bm{u}'$ and repeat the procedure.

After performing $N$ of these error-doubling experiments, each giving one set of doubling times $T^{i}(\delta_n)$, we compute the average over this ensemble
\begin{equation}
\langle T(\delta_n) \rangle_e = {1 \over N} \sum_{i=1}^{N} T^{i}(\delta_n)\,,
\label{eq:app9}
\end{equation}
which gives the FSLE (\ref{eq:fsle-def}) as
\begin{equation}
\lambda(\delta_n) = \frac{1}{\langle T(\delta_n) \rangle_e} \ln r \ .
\label{eq:app10}
\end{equation}
This procedure was used to compute the FSLE for the particle dispersion in the Kraichnan flow, i.e., to generate Fig.~\ref{fig:richardson}. We remark that the ensemble average of the doubling times corresponds to the time average of the inverse doubling times, which is the natural definition of the FSLE. To realize this write \cite{aurell1997predictability}
\begin{eqnarray}
\left\langle \frac{1}{T(\delta)} \!\right\rangle \!\equiv\! \frac{1}{T}\!\! \int_0^T\!\!\!\!\! \frac{dt}{T(\delta)}\!=\!  \frac{\sum_i \frac{1}{T^{i}(\delta)} T^{i}(\delta)}{\sum_i T^{i}(\delta)}\!=\!\frac{1}{\langle T(\delta) \rangle_e}\,.
\label{eq:app11}
\end{eqnarray}

In principle, it is possible to remove the threshold condition and compute the FSLE in terms of the average growth rate at a given (short) time interval $\Delta t$.  The idea is that after every $\Delta t$ the ratio $\delta(t+\Delta t)/\delta(t)$ is computed while the perturbed trajectory ${\bm u}'$ is rescaled to the original distance $\delta(t)$ keeping the same direction ${\bm u}'-{\bf u}$ in the phase space.  While this procedure is not a problem in the case of infinitesimal perturbations, for finite perturbations the rescaling can potentially displace ${\bm u}'$ out of the attractor. For this reason, the time interval $\Delta t$ should be as short as possible, typically the time step of the numerical integration.  After many realizations of this one-step experiment, the FSLE is defined as
\begin{equation}
\lambda(\delta) = \frac{1}{\Delta t} \left\langle \ln \left({\delta(t+\Delta t) \over \delta(t)} \right) \right\rangle
\label{eq:app12}
\end{equation}
This is the way we used to compute the FSLE for the shell model, i.e., to generate Fig.~\ref{fig:fsle}.

%%%%%%%%%%%%%%%%%%%%%%%%%%%%%%%%%%%%%%%%%%%%
\bibliography{biblio}

\end{document}